\documentclass[11pt]{article}

\usepackage{amsmath}
\usepackage{amsfonts}
\usepackage{mathrsfs}
\usepackage{xspace}
\usepackage{datetime}
\usepackage{eso-pic}
\usepackage{cite}
\usepackage{lmodern}
\usepackage[T1]{fontenc}
\usepackage[utf8]{inputenc}

\numberwithin{equation}{section}
\makeatletter
\renewcommand{\@seccntformat}[1]{%
  \csname the#1\endcsname.\ }
\makeatother

\usepackage{braket}
\usepackage[pdfusetitle,linktoc=all,colorlinks,allcolors=blue!75!black]{hyperref}
\usepackage{cleveref}

\def\p{\partial}
\def\O{\mathcal{O}}
\def\L{\mathcal{L}}
\def\a{\alpha}
\def\b{\beta}
\def\g{\gamma}
\def\r{\rightarrow}
\def\F{\mathcal{F}}
\def\G{\Gamma}
\def\H{\mathcal{H}}
\def\P{\mathcal{P}}
\def\K{\mathcal{K}}

\def\d{\delta}
\def\s{\sigma}
\def\ts{\tilde{\sigma}}
\def\e{\epsilon}

\def\D{\Delta}
\def\X{\mathcal{X}}
\def\P{\mathcal{P}}
\def\Dmu{\mathcal{D}_\mu}

\newcommand{\tL}{{\tilde L}}
\newcommand{\tbL}{{\tilde{\bar L}}}
\newcommand{\TT}{\texorpdfstring{\ensuremath{{T \bar T}}}{TTbar}\xspace}

\newcommand{\cK}{\mathcal K}
\newcommand{\cX}{\mathcal X}

\newcommand{\be}{\begin{equation}}
\newcommand{\ee}{\end{equation}}
\newcommand{\beal}{\begin{aligned}}
\newcommand{\bea}{\begin{eqnarray}}
\newcommand{\eea}{\end{eqnarray}}
\newcommand{\eeal}{\end{aligned}}
\newcommand{\bi}{\begin{itemize}}
\newcommand{\ei}{\end{itemize}}

\DeclareMathOperator{\tr}{tr}

\title{Classical and quantum symmetries of \TT\ - deformed CFTs\texorpdfstring{\vspace{5mm}}{}}

\author{
\texorpdfstring{
Monica Guica$^\S$, Ruben Monten${}^\dag$ and Ioannis Tsiares$^\S$ \vspace{1mm} \\
\\\vspace{1mm}
${}^\S$\emph{\small Universit\'e Paris-Saclay, CNRS, CEA,} 
\emph{\small Institut de Physique Th\'eorique, 91191 Gif-sur-Yvette, France} \\
${}^\dag$\emph{\small Mani L. Bhaumik Institute for Theoretical Physics,} \\
\emph{\small Department of Physics and Astronomy, University of California, Los Angeles, CA 90095, USA}
}{Monica Guica, Ruben Monten and Ioannis Tsiares}
}

\date{}

\begin{document}

\maketitle

%

\begin{abstract}
\vskip 3mm
\noindent
It has previously been proven that $T\bar T$ - deformed CFTs possess Virasoro $\times$ Virasoro symmetry at the full quantum level, whose generators  are obtained by simply transporting the original CFT  generators   along the $T\bar T$ flow. In this article, we explicitly solve the corresponding flow equation  in the classical limit, obtaining an infinite set of conserved charges whose action on phase space is well-defined even when the theory is on a compact space. The field-dependent coordinates that are characteristic of the $T\bar T$ deformation are shown to  emerge unambiguously from the flow. Translating  the symmetry  transformations from the canonical to   the covariant formalism, we find that they are different 
from those that
have been previously 
proposed in the Lagrangian context, but that they agree precisely with those obtained from holography. 
 We also comment on different possible bases for the symmetry generators and on how our explicit classical results could be extended perturbatively to the quantum level.
\end{abstract}

\tableofcontents

\section{Introduction}

The $T\bar T$ deformation \cite{
Smirnov:2016lqw,Cavaglia:2016oda} is a solvable irrelevant deformation of two-dimensional QFTs, which is remarkable for a number of different reasons. First, the deformation leads to a UV-complete theory \cite{Dubovsky:2013ira}, but which is non-local at the scale set by the irrelevant deformation parameter \cite{Dubovsky:2012wk}. Second, this deformation has found applications to many different areas of theoretical physics, such as integrable QFT bootstrap \cite{Smirnov:2016lqw}, the bosonic and the effective string in QCD \cite{Dubovsky:2012sh,Dubovsky:2013gi,Caselle:2013dra,Dubovsky:2016cog,Chen:2018keo}, phenomenology \cite{Dubovsky:2013ira} as well as non-AdS holography, either in the standard sense of a different asymptotic background metric \cite{Giveon:2017nie}, or as a change in the asymptotic boundary conditions for an AdS one \cite{Guica:2019nzm,McGough:2016lol}. Finally, the deformation is entirely solvable, as many observables have been computed exactly, such as the finite-size spectrum and the S-matrix\cite{
Smirnov:2016lqw,Cavaglia:2016oda,Dubovsky:2017cnj,Dubovsky:2018bmo}, as well as correlation functions \cite{Cardy:2019qao,Rosenhaus:2019utc,Kruthoff:2020hsi,Ebert:2022cle,Kraus:2022mnu},  to a certain extent. In all these cases, the deformation of the given observable is universally determined by the corresponding quantity in the undeformed QFT.

Of course, one of the most basic properties of a theory are its symmetries; those of $T\bar T$ - deformed \emph{CFTs}, i.e., when the seed theory already possesses Virasoro $\times$ Virasoro symmetry, deserve particular attention. The reason is that these infinite symmetries are preserved by the deformation, despite  the theory  becoming explicitly non-local. This fact is interesting both from the perspective of $T\bar T$ - deformed CFTs, and as a proof of concept that Virasoro symmetries need not be associated to  \emph{local} two-dimensional CFTs. Non-local UV-complete QFTs with Virasoro symmetry have been termed ``non-local CFTs'' in \cite{Guica:2021pzy}, where it was also shown how their non-locality could be reconciled with the Virasoro symmetry. Such non-local CFTs are expected to be quite relevant,  in the least,   to non-AdS holography, for example to the Kerr/CFT correspondence \cite{El-Showk:2011euy} and possibly also to flat space holography, as suggested by the recent discovery 
of an infinite set of $T\bar T$ - like symmetries in three-dimensional asymptotically flat spacetimes with a linear dilaton \cite{Georgescu:2022iyx}.

The first hint that $T\bar T$ - deformed CFTs may possess a doubly-infinite set of Virasoro symmetries came from the holographic analysis of \cite{Guica:2019nzm}, who showed that the holographic dual to $T\bar T$ - deformed CFTs with a large central charge is AdS$_3$ gravity with mixed boundary conditions for the  metric, and worked out the associated asymptotic symmetries. The asymptotic symmetry generators were found to be parametrized by two arbitrary functions of certain ``field-dependent'' coordinates that, as emphasized in \cite{Dubovsky:2012wk,Dubovsky:2017cnj}, capture much of the dynamics of $T\bar T$ - deformed CFTs. It was also shown that the generators corresponding to the natural Fourier basis for these functions \emph{rescaled} by the field-dependent radius of the coordinates satisfy a Virasoro $\times$ Virasoro algebra, with the same central charge as the undeformed CFT. This analysis was further perfected in \cite{Georgescu:2022iyx}, which resolved certain subtleties related to coordinate periodicities and also showed how to obtain the non-linear algebra of the ``unrescaled'' generators directly from the covariant phase space formalism.


The infinite set of symmetries uncovered by the holographic analysis were later confirmed to  exist also from the perspective of   classical field theory\cite{Guica:2020uhm},  through both a Lagrangian  and a Hamiltonian analysis. More precisely, it was shown that the classical action of a $T\bar T$ - deformed CFT was invariant under a certain set of field-dependent coordinate transformations, whose form closely resembled, but was however not identical to, that of the asymptotic symmetry generators  in the holographic analysis. The algebra of the associated conserved charges was computed using the Hamiltonian formalism; however, while one clearly found two commuting copies of the Witt algebra in infinite volume
, for $T\bar T$ - deformed CFTs on a compact space  the symmetry algebra was not entirely fixed, due to an ambiguity  in the zero mode of the field-dependent coordinate.


The full quantum proof of the existence of Virasoro symmetries in $T\bar T$ - deformed CFTs was given in \cite{Guica:2021pzy}. It simply relies on the fact that when placed on a circle,  $T\bar T$ - deformed CFTs should possess a well-defined quantum operator, herein denoted as  $\hat \X_{T\bar T}$, that drives the flow of the energy eigenstates. Then, the quantum operators defined via

\be
\p_\mu \tilde L_n = [\hat \X_{T\bar T}, \tilde L_n] \;, \;\;\;\;\;\;\;\;  \left. \tilde L_n \right|_{\mu=0} = L_n^{\mbox{\tiny{CFT}}} \label{fleqlt}
\ee 
are well-defined, satisfy by construction  a Virasoro algebra with the same central extension as that of the undeformed CFT and, moreover, are conserved.  The latter fact, which is the only one that does not follow trivially from our definition of the flowed generators, is a consequence  of the universality of the spectrum of $T\bar T$ - deformed CFTs \cite{LeFloch:2019rut,Guica:2021pzy}. This rather formal definition  of the extended symmetries of $T\bar T$ - deformed CFTs is  the most robust one that has been given to date. 

In this article, we revisit the symmetries of classical $T\bar T$ - deformed CFTs 
 in compact space by taking the classical limit of the quantum generators defined above, as the latter are guaranteed to have a well-defined action on the Hilbert space.   The corresponding classical Hamiltonian generators are derived by explicitly solving the flow equation \eqref{fleqlt} in the classical limit. 
  The field-dependent coordinates that are inherent to $T\bar T$ - deformed CFTs turn out to simply emerge from the flow equation, which also fixes the ambiguity in their zero mode.  The algebra of the resulting flowed generators is therefore completely unambiguous, and is given as expected by two commuting copies of the Witt algebra. The  flowed generators turn out to  precisely correspond to rescaled versions - by the field-dependent radii - of the ones in the natural Fourier basis,  in perfect agreement with the holographic results. The algebra of the unrescaled generators is accordingly given by a non-linear  modification of the Witt algebra.


We then translate the  symmetries of $T\bar T$ - deformed CFTs generated by these charges to Lagrangian language. We find that the resulting transformations are  distinct from the previously proposed symmetries \cite{Guica:2020uhm}  in the Lagrangian context,
in two qualitatively different ways: first, the field-dependent coordinate transformation is different, and as a result it now needs to be accompanied by a Weyl rescaling of the metric in order for the action to be  invariant; second, a rescaling of space and time by an amount proportional to the conserved charges of the background is necessary in order for the  transformation of the fields to be consistent with momentum quantization. These symmetries are identical to those described in the recent holographic analysis of \cite{Georgescu:2022iyx}. We thus obtain a fully consistent classical picture for the extended $T\bar T$ symmetries, by showing they agree across  all approaches that have been attempted.  


Finally, we discuss how to extend our explicit classical results to the quantum case. The main challenge is that the $T\bar T$ operator 
is not explicitly known at the quantum level. We discuss a perturbative algorithm that could fix it, at least in part. We also comment on a possible quantum definition of the ``unrescaled'' generators that classically correspond to the natural Fourier basis of symmetry generators.

%

This article is organised as follows. In section \ref{sec:classHam}, we review the Hamiltonian formulation of classical $T\bar T$ - deformed CFTs, following \cite{Jorjadze:2020ili}, and expand upon the methods of \cite{Kruthoff:2020hsi} to derive the explicit expression for the flow operator in the classical limit. We then solve the resulting classical flow equation for the infinite set of conserved charges, display the emergence of the field-dependent coordinate, and discuss their conservation and algebra.  In section \ref{sec:classLagr}, we translate our Hamiltonian results to Lagrangian language. Finally, in section \ref{sec:Quant}, we discuss several quantum aspects of the symmetries, such as the perturbative quantum definition of the flow operator and the possibility of having a fully quantum definition of the unscaled symmetry generators. 
The appendix contains a list of  Poisson brackets that have been used to  derive many results presented in the main text.

\section{Conserved charges from the classical limit of the \TT flow}
\label{sec:classHam}

We start by reviewing the rather remarkable fact \cite{Frolov:2019xzi,Jorjadze:2020ili} that, classically, the Hamiltonian density of a $T\bar T$ - deformed CFT is given by an extremely simple and universal formula. We will be following a slightly rewritten version of \cite{Jorjadze:2020ili}.

\subsection{Hamiltonian formulation of classical \TT\ - deformed CFTs \label{hamform}}

In the Hamiltonian formalism, the \TT deformation is defined as 

\be
\label{Hflow}
\p_\mu \H = - \O^{cls}_{T\bar T} = - (T_{tt} T_{\s\s} - T_{t\s}^2)
\ee
where we denote the stress tensor components $T_{tt}$ and $T_{t\s}$ by the standard notation $\H, \P$ and, in a classical field theory, the $T_{t\s}$ and  $T_{\s\s}$ component are  determined by the canonical variables and derivatives of the Hamiltonian density with respect to them. This information is more elegantly captured by the equal-time commutation relations of $\H,\P$ which, for an unconstrained system with an action that depends only on first time derivatives, take the form

\be
\{\H (\s),  \P (\tilde \s)\} = ( \H(\tilde \s) + T_{\s\s} (\s)) \,\d'(\s-\tilde \s)\ee
 
\be \{\H(\s), \H (\tilde  \s)\} = \{\P(\s), \P (\tilde \s)\} = (\P(\s)+ \P (\tilde\s)) \, \d'(\s-\tilde \s) \label{pbcurr}
\ee
 Note these commutators imply the conservation equations upon integration over the spatial coordinate, $\s$. 
In a CFT, $T_{\s\s} = \H$ by the vanishing of the trace of the stress tensor, but in a general classical field theory this will not be the case. 

The authors of \cite{Jorjadze:2020ili} derived the deformed Hamiltonian density from the flow equation \eqref{Hflow}, finding the universal expression \eqref{eq:HFlow} that only depends on the original Hamiltonian and momentum density, while  $\P$ remains undeformed. Alternatively, given the universal structure of the $\TT$ deformation, 
one may start by assuming that the deformed  Hamiltonian density should be just a function of the undeformed Hamiltonian and momentum density. By itself, this is an extremely strong assumption that, when plugged into the $\{ H (\s),  \H (\tilde \s)\}$ Poisson bracket, is sufficient 
%
 to
 completely fix the deformed Hamiltonian density in terms of the undeformed one
, irrespectively of the flow equation \eqref{Hflow}
\be
\label{eq:absol}
\H = \pm \sqrt{\P^2 + 4 a \H^{(0)} + 4 a^2} - b
\ee
where $a,b$ are integration constants. We will choose the positive branch
and let $b = 2 a$ to ensure $\H$ reduces to $\H^{(0)}$ in the $a \r \infty$ limit. We then use the $\{ \H (\s),  \P (\tilde \s)\}$ commutator to determine $T_{\s\s}$. Promoting $a,b$ to functions of $\mu$, we  find that the $T\bar T$ flow equation is entirely consistent with this solution, provided $2a=b= 1/2\mu$. The exact classical solution for the Hamiltonian density of a $T\bar T$ - deformed CFT is then

\be
\label{eq:HFlow}
\H = \frac{1}{2\mu} \left(\sqrt{1+ 4 \mu \H^{(0)} + 4 \mu^2 \P^2} -1\right)
\ee
irrespectively of the form of the Hamiltonian of the original CFT that is being deformed. 

It follows from  this exact expression  that the commutation relations  of $\H, \P$ are entirely determined by those of $\H^{(0)}, \P$ and are universal,  fact that has been  exploited to a large extent in \cite{Guica:2020uhm}. In particular, we can use the commutator \eqref{pbcurr} to show that  

\be
T_{\s\s}
= \frac{\H+ 2 \mu \P^2}{1+2\mu \H}
\ee
This implies that classical $T\bar T$-deformed CFTs satisfy the trace relation

\be
\label{eq:trEq}
\tr \, T = T_{\s\s} - T_{tt} = - 2 \mu \, \frac{\H^2-\P^2}{1+2\mu \H} = - 2 \mu \, \O_{T\bar T}^{cls}
\ee
It is interesting to ask whether this relation survives quantum-mechanically. One may use the fact that $\mu$ is the only scale in the deformed CFT to argue that the trace relation must still hold when integrated over space-time \cite{McGough:2016lol,Kraus:2018xrn}, 
%
i.e.  the local relation above is valid up to at most total derivative terms. 
Our quantum analysis in section \ref{sec:Quant} shows these terms are not present at least to first perturbative order in   the $T\bar T$ parameter.

\subsection{The flow operator and its classical limit \label{flopcls}}

As mentioned in the introduction, the Virasoro symmetries of $T\bar T$ - deformed CFTs can be most rigorously understood by studying the flow of the original CFT symmetry generators  under the operator that drives the flow of the energy eigenstates of $T\bar T$ - deformed CFTs on the cylinder

\be
\p_\mu |n_\mu\rangle = \hat\X_{T\bar T} |n_\mu\rangle
\ee
As described in \cite{Kruthoff:2020hsi}, this operator can be obtained from standard first order quantum-mechanical perturbation theory, and 
can be related to the change in the $T\bar T$ Hamiltonian as

\be
\p_\mu H = D - [H,\hat \X_{T\bar T}] \label{defXttb}
\ee
where $D$ are the diagonal elements of $\p_\mu H$ in the energy eigenbasis. Thus, in order to recover $\hat \X_{T\bar T}$, we need to understand how to decompose $\p_\mu H$ into a diagonal piece in the energy eigenbasis and an off-diagonal one, which can be further written as a total commutator with $H$. 

For this, remember the original definition  of the  $T\bar T$ operator  \cite{Zamolodchikov:2004ce} as an antisymmetric bilinear in the stress tensor components regulated via point splitting

\be
\O_{T\bar T} (x) =  \lim_{\e \r 0} \frac{1}{2}\,  \e^{\a\b} \e^{\g\d} T_{\a\g} (x+\e) \, T_{\b\d} (x) + \ldots \label{psplit}
\ee
where the $\ldots$ stand for total derivative terms. Such terms drop out when integrating the $T\bar T$ operator over the entire spacetime, as is the case when adding the deformation to the action; however, here we would like to add this operator to the Hamiltonian, which is only integrated over space at a fixed time. Consequently, the change in the Hamiltonian induced by the $\TT$ operator above takes the form
%
\begin{align}
\label{eq:HPflow}
	\p_\mu H &= -\int d\s \, \left(\O_\TT + \frac{d}{dt} V\right) 
\end{align}
where now $\O_{T\bar T}$ denotes only the point-splitted current bilinear in \eqref{psplit}, which is potentially divergent, and $V$ (which is also potentially divergent) should be chosen such that the resulting $\p_\mu H$ is a well-defined operator. Writing the time derivative of $V$ as a commutator with the Hamiltonian, this term can be absorbed by a shift in $\hat \X_{T\bar T}$, using \eqref{defXttb}. 

We would now like to compute the contribution of the $T\bar T$ operator itself to $\hat \X_{T\bar T}$, which involves splitting it  into a “diagonal” and  an “off-diagonal” part, where the latter is  written as a commutator with the Hamiltonian. For this, we  follow the manipulations  of \cite{Kruthoff:2020hsi}, taking into account operator ordering to the largest extent possible
%
\begin{align}
	\int d \s \, \O_\TT
	&= \frac12 \int d\s \, d \ts \left( \frac1R + G'(\s - \ts) \right) \e^{ab} \e^{cd} T_{ac} (\s) T_{bd} (\tilde \s)
	\nonumber \\
	&= \frac{H K + K H - 2P^2}{2R} + \frac12 \int d\s \, d\ts \, G(\s - \ts) \e^{cd} \left( T_{tc} \, \p_{\ts} \tilde T_{\s d} + \p_\s T_{\s c} \, \tilde T_{td} \right)
	\ \nonumber \\
	&= \frac{H K + K H - 2P^2}{2R} + \left[ H, \frac{i}2  \int d\s \, d\ts \, G(\s - \ts) (\H \tilde \P - \P \tilde \H) \right]
\end{align}
where $K \equiv \int d\s T_{\s\s}$ and  $G(\s)$ is the Green's function of the spatial derivative on the cylinder of circumference $R$
\begin{align}
\label{eq:GDef}
	G(\s) = \frac12 \text{sign}(\s) - \frac{\s}R
	\ , \qquad
	G'(\s) = \d(\s) - \frac1R
\end{align}
In the last step, we used a Hamiltonian form of the conservation equation, i.e. the commutators $[H, \H] = - i \p_\s \P$ and $[H, \P] = - i \p_\s T_{\s\s}$, which can be inferred from \cref{pbcurr} and are expected to hold at the full quantum level. The point-splitting parameter $\e$ is implicitly present through the ordering of the current operators.

According to our previous discussion, the integrated form of the trace equation \eqref{eq:trEq} is 
\be
K = H - 2 \mu \int \O_{T\bar T} + \frac{d W}{dt}
\ee
where the last term is a potential quantum ambiguity that is not present classically. It is not hard to see this term can again be absorbed into $\hat \X_{T\bar T}$, yielding a contribution proportional to $H W+ W H$. We are thus left with the following equation for $\int d\s \O_{T\bar T}$ or, almost equivalently, $\p_\mu H$

%
\be
\p_\mu H + \frac{\mu }{R} (H \, \p_\mu H + \p_\mu H \, H)= - \frac{H^2 - P^2}R  - \left[ H, \frac{i}2 \int d\s \, d\ts \, G(\s - \ts) (\H \tilde \P - \P \tilde \H) + \ldots\right] \label{symmfleq}
\ee
where the $\ldots$ stand for the ambiguities that we have discussed. This is an equation of the type $X +\mu H X +\mu X H = A$, whose solution  takes the form 

\be
X = \sum_{m,n}  \a_{m,n}\, \mu^{m+n} H^m \! A H^n 
\ee
where the coefficients $\a_{m,n}$ satisfy the recursion relation  $\a_{m,n} +\a_{m-1,n}+\a_{m,n-1} =0$. Taking $A$ to consist of the  $\int G \H \P$ term and the $\ldots$,  this formally yields the solution for the flow operator.

For most of this article, we will be interested in the classical limit of the flow equation.   Then, it is  easy to show that the associated classical flow operator defined through $\hat \X_{T\bar T} \r_{cls}  i \X_{T\bar T} $ is 

\be
\label{eq:XDef}
\X_\TT = \frac{R}{R_H} \int d\s \, d\ts \, G(\s - \ts) \H(\s) \P (\tilde  \s)\ , \;\;\;\;\;\; R_H\equiv R+2\mu H
\ee
up to  an undetermined phase space function whose Poisson bracket with the Hamiltonian vanishes and that we can freely set to zero.  The relationship between the classical $T \bar T$  and classical flow operator is
\be
\p_\mu H =-	\int d\s \, \O_\TT^{cls} = - \frac{H^2 - P^2}{R_H}+ \{H, \X_\TT\}  
\ee
Introducing the ``covariant derivative'' along the flow 

\be
\Dmu \equiv \p_\mu + \{ \X_\TT, \; \cdot \; \} 
\ee
the above can be written as a flow equation that the Hamiltonian satisfies

\be
\Dmu H = - \frac{H^2-P^2}{R_H} \label{flowham}
\ee
This equation holds in fact also quantum-mechanically, by replacing $\X_\TT$ by the corresponding operator, and Poisson brackets by commutators. We will denote the quantum version of $\Dmu$ by $\hat{\mathcal{D}}_\mu$.

One can easily check that the expression \eqref{eq:XDef} for $\X_{T\bar T}$ is consistent with expectations from the flow of the energy eigenstates on the cylinder. 
As argued in \cite{LeFloch:2019rut}, the universality of the $T\bar T$ - deformed spectrum implies that the zero modes of the flowed Virasoro operators \eqref{fleqlt} are related to the Hamiltonian and momentum operators in $T\bar T$ - deformed CFTs as 
\be
\widetilde L_0 + \widetilde{\bar L}_0   = H + \frac{\mu}{R} (H^2-P^2)  \;, \;\;\;\;\;\;\; \widetilde L_0 - \widetilde{\bar L}_0   = P
\ee
Since $\tilde L_0, \tilde{\bar L}_0$ satisfy the homogeneous flow equation \eqref{fleqlt}, we find
%
 precisely the quantum version of    the flow  equation \eqref{flowham} above, together with $ \hat{\mathcal{D}}_\mu P=0$. In fact, we could have alternatively defined $\X_\TT$ through the flow equations for the Virasoro zero modes that, as shown above, determine 
%
 the Poisson brackets of $\X_{T\bar T}$ with $H,P$ in the classical limit.  The solution is precisely \eqref{eq:XDef}, taking into account the fact that $\p_\mu H = \int d\s \p_\mu \H$, with $\H$  given classically by \eqref{eq:HFlow}. 
%
%
%
%
%
%
%
%

The analysis we presented assumed that the energy eigenstates of the seed theory were non-degenerate. However, this assumption is not true for two-dimensional CFTs, where the Virasoro symmetry leads to a high level of degeneracy. The states of the system then carry an additional label in addition to  the energy and momentum eigenvalues, for example the eigenvalue under the KdV charges. Taking this into account, 
%
the \TT deformation can  be treated using quantum-mechanical degenerate  perturbation theory 
 that does not get lifted at any order \cite{Sakurai:2011zz}. 

The flow of the KdV charges as the $T\bar T$ parameter is changed can be studied in a similar fashion to the flow of the Virasoro generators. Namely, one can define a flowed KdV generator via 

\be
\p_\mu \tilde P_s = [\X_{T\bar T}, \tilde P_s]
\ee
subject to the initial condition that at $\mu=0$, $\tilde P_s$ corresponds to the KdV generator in the undeformed CFT. Since in the original CFT, the $P_s$ have a known expression in terms of the Virasoro generators $L_n$, the above flow equation implies that the flowed $\tilde P_s$ will be given by the \emph{same}  expression, but now in terms of the flowed generators $\tilde L_n$. The flow equation also implies that the eigenvalue of $\tilde P_s$ in a flowed energy eigenstate is the same as the corresponding eigenvalue in the undeformed CFT. 

At a first glance, this statement appears to contradict 
%
the known fact \cite{LeFloch:2019wlf} that the expectation values of the  KdV charges in the $n$th energy eigenstate, $\braket{P_s}_n$, do flow nontrivially with $\mu$. The puzzle is resolved by noting that even the lowest KdV charges, namely the left- and right-moving energies, suffer from this problem: indeed, while  the flow equation correctly indicates that the expectation value of $\langle \tilde L_0 \rangle$ is unchanged under the flow, the physical energies of the system are related to this expectation value as 

\be
\braket{\tL_0} = H_L R_u \;, \;\;\;\;\;\;  \braket{\tbL_0}  = H_R R_v
\ee
where $H_{L,R} \equiv (H\pm P)/2$ are the left/right Hamiltonians and the quantities $R_{u,v}$ (which will turn out to correspond to the radii of the field-dependent coordinates, hence the notation) are for now defined as 

\be
\label{eq:RuvFlow}
R_u = R + 2\mu H_R \;, \;\;\;\;\;\;\; R_v = R+2\mu H_L
\ee
Since, using \eqref{flowham}, the quantities $R_{u,v}$ satisfy the following  flow equations 

\be
{\mathcal{D}}_\mu R_u=  \frac{2 H_R R_u}{R_H}\;, \;\;\;\;\;\;{\mathcal{D}}_\mu R_v = \frac{2 H_L R_v}{R_H} \label{flowRuv}
\ee
it follows that the physical energies will also satisfy non-trivial flow equations, as we already saw. The KdV charges work in the same way, as the ones  considered in \cite{LeFloch:2019wlf} differ by a factor of $R_u^s$ from the $\tilde P_s$ defined above. Since $R_u$ flows non-trivially, so does $\tilde P_s/ R_u^s$, and it can be explicit checked that the flow of these charges precisely agrees with that given in \cite{LeFloch:2019wlf}.  The advantage of recasting them in terms of flowed generators is that, at least classically, we  gain access to their explicit expression in terms of the flowed left/right-moving stress tensor, whose construction we  explain next.

\subsection{Explicit solution to the flow equation}
\label{sec:fieldDeptCoord}


We would now like to use the classical flow operator, $\mathcal{X}_{T\bar T}$, obtained in the previous subsection to solve explicitly  for the classical counterparts, denoted $\tilde Q_m$, of the quantum generators $\tilde L_m$, which   by definition
%
 satisfy the homogeneous flow equation $\mathcal{D}_\mu \tilde Q_m=0$. Writing 
 
 \be
 \widetilde Q_m = \int d\s e^{i m (\s+t)}  \mathscr{H}_L \;, \;\;\;\;\;\; \widetilde{\bar Q}_m = \int d\s e^{-i m (\s-t)}  \mathscr{H}_R \label{flchar}
 \ee
 our task is equivalent to solving the following homogeneous flow equations for the currents
\be\label{eq:homogfloweq}
\Dmu \mathscr{H}_{L,R} =0 \;, \;\;\;\;\;\;\;\; \left. \mathscr{H}_{L,R} \right|_{\mu=0} = \H^{(0)}_{L,R}
\ee
Note that charge conservation requires these currents to be (anti)chiral 
\be
 \frac{d}{dt} \mathscr{H}_{L,R} = \pm \p_\s \mathscr{H}_{L,R} \label{chircond}
 \ee
%
%
A natural starting point for  constructing an explicit solution to the equations (\ref{eq:homogfloweq}) is to consider the flow of the left/right Hamiltonian currents  $\H_{L,R} = (\H\pm \P)/2$, with $\H$ given in \eqref{eq:HFlow}. Setting $R=1$ for simplicity,  these read 
\bea
\mathcal{D}_\mu \H_L &=&\frac{1}{R_H} \p_\s \left[2\H_L\left(\chi_{R}+\frac{\mu\mathcal{X}_{T\bar{T}}}{R}\right)-\mu\partial_{\mu}\mathcal{H}\left(\chi_{\mathcal{P}}-\frac{2\mu\mathcal{X}_{T\bar{T}}}{R}\right)\right]  - \frac{2 H_R}{R_H} \H_L \nonumber \\[4pt]
\mathcal{D}_\mu \H_R &=&\frac{1}{R_H} \p_\s \left[2\H_R\left(\chi_{L}-\frac{\mu\mathcal{X}_{T\bar{T}}}{R}\right)+\mu\partial_{\mu}\mathcal{H}\left(\chi_{\mathcal{P}}-\frac{2\mu\mathcal{X}_{T\bar{T}}}{R}\right)\right]  - \frac{2 H_L}{R_H} \H_R 
\label{eq:HLRFlow}
\eea
where $\chi_\P \equiv \chi_L -\chi_R$ and the quantities $\chi_{L,R}$ are defined as
\be
\p_\s \chi_{L,R} \equiv \H_{L,R} - H_{L,R} \label{defchiLR}
\ee
which resembles the definitions of \cite{Guica:2020uhm}, except that the zero mode of the function has been explicitly subtracted. 
The solution for them can be written compactly as 

\be
\chi_{L,R} = \int d\tilde \s \, G(\s-\tilde \s) \H_{L,R} (\tilde \s)
\ee
One immediately notes, using the flow equations \eqref{flowRuv} for $R_{u,v}$, that the rescaled currents $R_u \H_L$ and $R_v \H_R$ satisfy flow equations whose right-hand-side is a total $\s$ derivative


\be
\label{eq:RuHLFlow}
\mathcal{D}_\mu \left(R_u \H_L\right) =  \p_\s \mathcal{A}_+\;, ~~~~~ \mathcal{D}_\mu \left(R_v \H_R\right) =  \p_\s \mathcal{A}_-
\ee
where $\mathcal{A}_\pm$ can be trivially read off from \eqref{eq:HLRFlow}. 
%
Introducing the notation

\be\label{eq:DuDv0}
\Delta \hat{u}\equiv \frac{2\mu}{R_u}\left(\chi_{R}+\mu\mathcal{X}_{T\bar{T}}\right)\;, \;\;\;\;\;\;
\Delta \hat{v}\equiv \frac{2\mu}{R_v}\left(\chi_{L}-\mu\mathcal{X}_{T\bar{T}}\right),
\ee
we find that the flow equations for these quantities are given by
\be
\Dmu \Delta\hat{u}=\frac{2}{R_uR_H}\left(\chi_{R}+\mu\mathcal{X}_{T\bar{T}}+\frac{\mu R_H}{R_v} \mathcal{A}_{-}\right)\;, ~~~~~\Dmu \Delta\hat{v}=\frac{2}{R_vR_H}\left(\chi_{L}-\mu\mathcal{X}_{T\bar{T}}+\frac{\mu R_H}{R_u} \mathcal{A}_{+}\right)
\ee
To understand how the flow equation is solved, it is useful to first study the $\mu \r 0$ limit of \eqref{eq:HLRFlow}, in which only the $2\p_\s (\H_L \chi_R)$ term survives on the right-hand-side of the first equation.  Since $\Dmu \Delta \hat u = 2 \chi_R$ in this limit, while $\D \hat u$ itself is $\O(\mu)$, it is clear that this $\O(\mu^0)$ term in the flow equation can be absorbed by a term that corresponds to the $\mu \r 0$ limit of $\p_\s (\H_L \D \hat u)$.  Inspired by this, we evaluate, now exactly in $\mu$

\be\label{eq:flowDuDv0}
\beal
&\Dmu \left(R_u\H_L(\Delta\hat{u})^k \right)=k(\Delta\hat{u})^{k-1}\mathcal{A}_++\partial_\s\left[\mathcal{A}_+(\Delta\hat{u})^k\right]
\eeal
\ee
An analogous equation holds for the right-movers. 
The solution to the flow equation \eqref{eq:homogfloweq} can now be iteratively constructed: starting from \eqref{eq:RuHLFlow}, one absorbs the term on the right-hand-side by subtracting $\p_\s (R_u \H_L \D \hat u)$ from the initial trial function. Using \eqref{eq:flowDuDv0} with $k=1$, we find
 
\be
\Dmu [R_u \H_L - \p_\s ( R_u \H_L \D \hat u)]= -\p_\s^2 [\mathcal{A_+} (\D \hat u)^2]
\ee
The term on the right-hand-side  of this equation can be absorbed by adding $\p_\s^2 (R_u \H_L (\D \hat u)^2)$ to the current, and so on. The full solution  to the problem (\ref{eq:homogfloweq}) is then, at least formally

\be
\mathscr{H}_L = R_u  \sum_{k=0}^\infty \frac{(-1)^k}{k!} \p_\s^k [\H_L (\Delta \hat{u})^k]\;, \;\;\;\;\; \mathscr{H}_R = R_v  \sum_{k=0}^\infty \frac{(-1)^k}{k!} \p_\s^k [\H_R (\Delta \hat{v})^k] \label{eq:solutionH}
\ee
 The discussion so far holds for functions on phase space at $t=0$. One can  check that  these currents do not quite satisfy the chirality conditions required by charge conservation,  obtained by replacing $\frac{d}{dt}$ by $\p_t - \{ H, \; \cdot \;\}$ in \eqref{chircond}, unless they acquire an explicit time dependence. This can be easily achieved by adding to $\D \hat u, \D \hat v$ a piece that is linear in time, so now

\be
\label{eq:DuDv}
\Delta \hat{u} =\frac{2\mu}{R_u}\left(\chi_{R}+ \mu\mathcal{X}_{T\bar{T}}\right)- \frac{4\mu H_R}{R_H}t\;, \;\;\;\;\;\;
\Delta \hat{v}= \frac{2\mu}{R_v}\left(\chi_{L}-\mu\mathcal{X}_{T\bar{T}}\right)+\frac{4\mu H_L}{R_H}t
\ee
One could have alternatively obtained this time dependence by taking into account the fact that the states at $t\neq 0$ differ by a factor of $e^{-i E t}$ from those at $t =0$, which results in a contribution proportional to $\p_\mu E \, t$ to the flow operator, as also explained in \cite{Guica:2020eab}. 

The conserved charges that satisfy the flow equation are simply given by the Fourier modes of the flowed currents, as in \eqref{flchar}. Plugging in the explicit expressions 
 (\ref{eq:solutionH}) and integrating by parts the total derivative terms, we find that the final result exponentiates into the following simple expressions  
\be
\widetilde{Q}_n=
R_u\int d\s e^{in\left(\s+t+\Delta\hat{u}\right)}
\H_L \;, \;\;\;\;\;\;\;
\widetilde{\bar{Q}}_n=
R_v\int d\s e^{-in\left(\s-t +\Delta\hat{v}\right)}\H_R \label{explflgen}
\ee
Thus, we note that the conserved charges can be rewritten in terms of the \emph{physical} Hamiltonian densities $\H_{L,R}$, provided the functions that parametrize these charges now depend on two field-dependent coordinates given by 

\be
\label{eq:uvDef}
u \equiv R_u (\s + t + \D \hat u ) \;, \;\;\;\;\;\;\; v \equiv R_v (\s-t+\D \hat v)
\ee
where the reason for pulling out the factors of $R_{u,v}$ is that this way

\be
\p_\s u = 1+2\mu \H_R \;, \;\;\;\;\; \p_\s v = 1+2 \mu \H_L
\ee
exactly as in \cite{Guica:2020uhm}, where we have used \eqref{eq:DuDv} and \eqref{defchiLR}.  From this perspective, the quantities $R_{u,v}$ are the field-dependent radii of these field-dependent coordinates, thus explaining our notation. 
The ratios $u/R_u$ and $v/R_v$ will be denoted henceforth as $\hat u, \hat v$.
 The time derivatives of these coordinates are given by 
\be
\label{eq:dtudsu}
\p_t u = \frac{R_u (1+2\mu P)}{R_H} \;, \;\;\;\;\;\;\; \frac{d u}{dt} =\p_t u - \{H, u\} = \frac{1+2\mu \P}{1+2\mu \H} \p_\s u
\ee
and similarly  for the right-movers

\be
\p_t v = -\frac{R_v (1 - 2\mu P)}{R_H} \;, \;\;\;\;\;\; \frac{dv}{dt}= \p_t v - \{H, v\}  = -\frac{1-2\mu \mathcal{P}}{1+2\mu \mathcal{H}} \p_\s v
\ee
These equations lead to the conservation of any charge of the form 
\eqref{eq:QDef} for arbitrary $m$.

It is interesting to note that the field-dependent coordinates \eqref{eq:uvDef} were not put in by hand; rather, they emerged from the flow equation for the currents. As a consequence, the expression for them is entirely unambiguous, and any Poisson bracket of these quantities can be explicitly computed using the expressions \eqref{eq:DuDv} and the Poisson brackets spelled out in the appendix.  In particular, the Poisson brackets of $\chi_{L,R}$ are entirely well-defined, as the zero mode of these quantities is zero by definition. There is therefore no obstruction to computing the charge algebra explicitly. 

This is, however, an extremely tedious computation, which is superseded by the fact that the algebra of the generators $\widetilde Q_m$ is Witt by construction.

\be
i \{ \widetilde Q_m , \widetilde Q_n \} =  (m-n)\, \widetilde Q_{m+n}
\ee
and similarly for the right-movers. Of course, the full quantum algebra is Virasoro, but our classical computation is, as usual, unable to capture the central extension.  We have also computed this algebra by brute force using the explicit expressions for the charges and confirmed this result. Even though the intermediate steps of the calculation are rather involved, 
%
 the various contributions nevertheless conspire in such a way that the final result is remarkably a Witt algebra, in agreement with our expectation from the flow equation. 

Note, from \eqref{explflgen}, that the flowed generators are proportional to an overall factor of $R_{u,v}$. It appears rather natural to define a set of  ``unrescaled'' pseudoconformal symmetry generators as

\be
\label{eq:QDef}
Q_m = \int d\s \, e^{i m \hat u} \, \H_L \;, \;\;\;\;\;
\bar{Q}_{m} = \int d\sigma \, e^{-i m \hat v}\, \mathcal{H}_R 
\ee
which corresponds to the natural Fourier basis for the functions that label the conserved charges. 
The algebra of these unrescaled generators is given by

%

\be
i \{ Q_m, Q_n \} = \frac{1}{R_u} (m-n) Q_{m+n} + \frac{4\mu^2 H_R}{  R_H R_u} (m-n) Q_m Q_n 
\nonumber
\ee

\be
i \{ \bar Q_m, \bar Q_n \} = \frac{1}{R_v} (m-n) \bar Q_{m+n} + \frac{4\mu^2 H_L}{ R_H R_v} (m-n)\bar  Q_m \bar  Q_n 
\nonumber
\ee

\be
i \{  Q_m, \bar Q_n \} = - \frac{2\mu (m-n)}{ R_H} Q_m \bar Q_n \label{eq:brackQmQnv2}
\ee
We note that the commutators of the left/right Hamiltonian $H_L=Q_0$, $H_R = \bar Q_0$ with the left conserved charges is given by 

\be
i \{ H_L, Q_n\} = -  \frac{n R_v}{R_H} \, Q_{n} \;, \;\;\;\;\;\;\;\;i\{ H_R, Q_n\} = \frac{2\mu n H_R}{R_H} \, Q_{n}  \label{clscommH}
\ee
and similarly for the right movers. The form of these Poisson brackets is very important, as they imply that $i \{ P, Q_n\} = -n Q_n$, and thus the commutation relations of the conserved charges are consistent with momentum quantization, as required if they are to have a well-defined action on the Hilbert space when the theory is in finite size. One can also check that these brackets precisely agree with the classical limit of the quantum commutators \eqref{qcomm}, first computed in \cite{Guica:2021pzy}.

\section{Lagrangian interpretation of the symmetries}
\label{sec:classLagr}

In the previous section, we have discussed how the classical limit of the quantum flow equation, which is guaranteed to produce generators that have a well-defined action on phase space, yields the classical conserved charges of $T\bar T$ - deformed CFTs in the Hamiltonian formalism, including an unambiguous definition of the field-dependent coordinates in this setting.  In this section, we would like to understand the action of the associated symmetries in the Lagrangian formalism, as this is a more natural language for describing space-time symmetries.

 This  can be readily obtained by  translating the action of the Hamiltonian generators on the fields to Lagrangian language. Interestingly, we find that the resulting symmetries differ from those proposed in \cite{Guica:2020uhm}, though they exactly agree  on-shell with the  asymptotic symmetry group generators of the spacetime holographically dual to $T\bar T$ - deformed CFTs \cite{Guica:2019nzm,Georgescu:2022iyx}.  In the following, we first review the relevant parts of these previous works, then compute the field transformations in the Hamiltonian formalism, and finally translate them to Lagrangian language, showing that the resulting transformations leave the action invariant.

\subsection{Brief review of previous results}

Let us start by reviewing the results of \cite{Guica:2020uhm}. We consider a classical $T\bar T$ - deformed CFT and work in null coordinates, $U,V = \s \pm t$.  In these coordinates, the trace relation - which is an off-shell property -  takes the form $T_{UV} = -2 \mu (T_{UU} T_{VV} - T_{UV}^2)$. Consequently,   the stress tensor of this theory only has two independent   components off-shell, which we parametrize by

\be
\mu \L \equiv - \frac{T_{UV}}{T_{VV}}
\;, \;\;\;\;\; \mu \bar \L\equiv - \frac{T_{VU}}{T_{UU}}
\ee
Given that under a coordinate transformation $x^\a \r x^\a +\xi^\a$

\be
\d S = - \int d^2 x \, T^{\a}{}_{\b}\, \p_\a \xi^\b \label{varactdiff}
\ee
one can easily show that the action is invariant under
 the  following field-dependent coordinate transformations 
\be
U \r U + \e f(u) \;, \;\;\;\;\;\;  V \r V - \e \bar f(v) \label{olddiff}
\ee 
provided we choose the ratios of the derivatives of the field-dependent coordinates to be the same as that of the stress tensor components

\be
\p_V u = \mu \bar \L \p_U u\;, \;\;\;\;\; \p_U v = \mu \L \p_V v
\ee
These equations were shown to always  have a solution \emph{off-shell} for $u,v$, which is important to show invariance of the action. On-shell,  $\L, \bar \L$ become functions of only one of the field-dependent coordinates, namely $\L= \L(u)$ and $\bar \L = \bar \L(v)$, and it is moreover possible to choose the gauge $\p_u U = \p_v V =1$, where $\p_x X$ is the inverse matrix to $\p_X x$. The components of this matrix satisfy

\be
 \p_v U = - \mu \bar \L \p_u U\;, \;\;\;\;\;\;\;\;\p_u V = - \mu \L \p_v V
 \ee
which, together with the previous gauge condition results in the following implicit  on-shell solution for the field-dependent coordinates $u,v$ in terms of the fixed ones and the on-shell stress tensor components

\be
U = u - \mu \int^v \! \bar \L(v') dv' \;, \;\;\;\;\;\;\; V = v - \mu \int^u \! \L(u') du'
\ee
The Noether charges associated with these symmetries take the form

\be
Q_f = \int d\s f(\hat u)\, \H_L   \;, \;\;\;\;\;\; \bar Q_{\bar f} = \int d\s \, \bar f(\hat v) \H_R \label{conschlagr}
\ee
Due to the periodicity constraint, $f, \bar f$ only depend on the rescaled field-dependent coordinates, which have fixed periodicity, $1$. The integrands  $\H_{L,R}$ denote the stress tensor components $T_{tU}$ and, respectively, $-T_{t V}$, whose expression in terms of $\L, \bar \L$ reads 

\be
\H_L = \frac{\L (1+\mu \bar \L)}{2(1-\mu^2 \L \bar \L)} = \frac12 \L \p_\s u \;, \;\;\;\;\;\; \H_R = \frac{\bar \L (1+\mu \L)}{2(1-\mu^2 \L \bar \L)} = \frac12 \bar \L \p_\s v \label{HLRlagr}
\ee
where the second set of equalities only holds on-shell.  The left and right-moving global energies $H_{L,R}$ are simply given by the choice $f=1$ and, respectively, $\bar f =1$. 

All this can be made very explicit for the case of the $T\bar T$ - deformed free boson, whose action is 

\be
S= \frac{1}{4\mu} \int dU dV \left(1 - \sqrt{1+ 8\mu \p_U \phi \p_V \phi} \right)
\ee
The stress tensor can be computed explicitly, yielding explicit formulae for $\L, \bar \L$, and can be checked to satisfy the off-shell relation $tr T =-2\mu \O_{T\bar T}$. Under the above transformations, the variation of the basic field is 
\be
\d_f \phi = - f(u) \p_U \phi \label{fdeptransl}
\ee 
One may use this to compute the variation of the conserved charges, for example $H_L$, under the diffeomorphism \eqref{olddiff}. Generally speaking, this variation is

\be
\d_\xi H_L = \int d\s \left(\p_\s \xi^U T_{UU} -\p_\s \xi^V T_{VU} - \frac{\p \H_L}{\p(\p_\b \phi)} \p_\b \xi^\a  \p_\a \phi\right)
\ee
using the conservation of the stress tensor. For a left-moving diffeomorphism of the form \eqref{olddiff}, we find
\be
\d_f H_L =-\frac{1}{2} \int d\s \frac{\L \p_\s f(\hat u)}{(1+ \mu \sqrt{\L \bar \L})^2}
\ee
which, strangely enough, does not yield a conserved charge of the form \eqref{conschlagr}, as one would have expected from the assumed existence of a charge algebra. 
 
\bigskip

Let us now also briefly summarize the relevant results from the holographic analyses of \cite{Guica:2019nzm} and especially \cite{Georgescu:2022iyx}. The holographic dual to $T\bar T$ - deformed holographic CFTs is AdS$_3$ with mixed boundary conditions for the metric. The radial gauge diffeomorphisms that leave invariant these boundary conditions are given by

\be
\xi^U = f(u)+\mu \int^v\! \bar \L \bar f'(v)\;, \;\;\;\;\;\;
\xi^V = \bar f(v)+\mu \int^u \! \L f'(u)  \label{xiV} \;, \;\;\;\;\;\; \xi^\rho = \rho \,(f'(u)+\bar f'(v))
\ee
where we have neglected the terms proportional to the holographic anomaly. These diffeomorphisms are on-shell, so $\L=\L(u)$ etc.  Quite importantly, the functions $f, \bar f$ must carry winding (linear in the respective coordinate) in order for these coordinate transformations to be well-defined, namely to not change the periodicity of the fixed $T\bar T$ coordinates $U,V$. Nevertheless, the conserved charges associated to these diffeomorphisms  only depend on the  periodic part of the function $f$, as required by charge conservation. The winding is entirely determined by the conserved charges associated with the periodic part of the function and, in our conventions, is given by\footnote{Note that the ${}'$ on $f$ is a $u$ derivative, which thus brings down a factor of $2\pi/R_u$. } 

\be
w_f =  \frac{2\mu R_v}{ R_H} \bar Q_{\bar f'} + \frac{4\mu^2 H_R}{  R_H} Q_{f'} \;, \;\;\;\;\; w_{\bar f} = - \frac{2\mu R_u}{ R_H}  Q_{ f'} - \frac{4\mu^2 H_L}{ R_H} \bar Q_{\bar f'} \label{windingffb}
\ee
The winding contribution to the asymptotic diffeomorphism has been referred to as a ``compensating diffeomorphism'' in \cite{Georgescu:2022iyx}, in the sense that it compensates the unphysical winding of the field-dependent coordinates that is associated with the periodic part of the functions. This compensating diffeomorphism takes the simple form 

\be
\xi_c^U = (w_f+w_{\bar f}) u - w_{\bar f} U 
\;, \;\;\;\;\;\;
\xi_c^V = (w_f+w_{\bar f}) v - w_f V
\ee
Notice that the proposed symmetry transformations in the Lagrangian and holographic settings are different. For example, for a purely left-moving transformation holography predicts

\be
\xi^U = f(u) - \frac{2\mu Q_{f'}}{R_H}(  u -  R_u U) \;, \;\;\;\; \xi^V = \mu \int^u \!\L f' - \frac{2\mu Q_{f'}}{R_H} (v+ 2\mu H_R V) \;, \;\;\;\; \xi^\rho = \rho \left(f'- \frac{2\mu Q_{f'}}{R_H}  \right) \label{holodiffs}
\ee
which is rather different from \eqref{olddiff}, and not only in what concerns the compensating diffeomorphisms. In particular, one notes that for the above diffeomorphism, $\p_\a \xi^\b T^{\a}{}_{\b} \neq 0$. In the following, we would like to understand which  Lagrangian symmetries to consider, by making a comparison  with the results of the Hamiltonian analysis.

\subsection{Symmetry variations in the Hamiltonian formalism}

The starting point of the Hamiltonian analysis are the conserved charges, which are given by  

\be
Q_f = \int d\s f(u) \H_L 
\ee
where $\H_L$ is  again a function of the canonical variables, as in section \ref{sec:classHam}. For simplicity, we will be concentrating on the left-movers only. The transformations of the basic fields in the theory are given by 
\be
\d_f \phi = \{ Q_f, \phi\}  \;, \;\;\;\;\;\;\; \d_f \pi = \{ Q_f, \pi\}
\ee
where we will be considering the action of the symmetries on a single boson; the generalisation to several fields is trivial. Translating the right-hand side to the Lagrangian formalism using $\dot \phi = \p_\pi \H$ points towards the transformation that we should be using.  To compute the above field variations, we use the standard commutation relations

\be
\{ \H, \tilde \phi\} = - \p_\pi \H(\s) \, \d'(\s-\tilde \s) \;, \;\;\;\;\;\;\;\{ \P, \tilde \phi\} = - \phi'(\s) \, \d'(\s-\tilde \s)
\ee
where  the second term is a consequence of $\P = \pi \phi'$. All the Poisson brackets that follow can be derived from these basic commutators.  Let us first compute $\d_f\phi$ 

\be \{ Q_f, \phi\} = -  f(\hat u)\, \frac{ \phi'+\p_\pi \H }{2} + \int d \tilde \s \hat f'( \hat{\tilde u} )\tilde \H_L \left \{ \frac{\tilde u}{R_u},\phi \right\} 
\ee
where a hat on $f'$ denotes the fact that the derivative is taken with respect to $\hat u$, rather than $u$. We thus have $f' = \hat f' /R_u$, an important relation to keep in mind.

Using  $ \dot \phi = \p_\pi \H$, the term in the first parenthesis yields $-f  \p_U \phi$, so this is nothing but the expected field-dependent translation \eqref{fdeptransl}. The  commutator of the field with the field-dependent coordinate yields an additional contribution.  To compute this correction, we first estimate 
%

\be
\left\{ \frac{\tilde u}{R_u}, \phi \right\} = \frac{2\mu \Delta \hat u}{R_H} \, \frac{\phi'+\p_\pi \H}{2}  + \frac{2\mu}{R_u} \left( G(\tilde \s -\s) + \frac{R_v \Delta \hat v}{R_H} - \Delta \hat{\tilde u} -2t\right) \, \frac{\phi'-\p_\pi \H}{2} 
\ee
where $\D \hat u$ and $\D \hat v$ have been defined in \eqref{eq:DuDv}.
%
Performing the integral, we find  

\be
\d_f \phi = - \left(  f(\hat u) - \frac{2\mu \D \hat u}{R_H} Q_{\hat f'}\right)\, \frac{\phi'+\p_\pi \H}{2} 
 + \frac{2\mu}{R_u} \left[ \F_f + Q_{\hat f'} \left( \frac{R_v \Delta \hat v}{R_H} -2t\right)  \right]\, \frac{\phi'-\p_\pi \H}{2} \label{varphi}
 \ee
where we have defined

\be
\F_f (\s) \equiv \int d\tilde \s f'(\hat {\tilde u}) \tilde \H_L  [G(\tilde \s -\s) - \Delta \hat{\tilde u}] \label{Ff}
\ee
We may also compute the Poisson bracket with the conjugate momentum
\bea
\{ Q_f, \pi\}& = & - \p_\s \left( f(\hat u) \frac{\pi + \p_{\phi'} \H}{2} \right) + \int d\tilde \s \tilde \H_L \hat f'(\hat{\tilde u}) \left\{ \frac{\tilde u}{R_u}, \pi \right\}  \label{varpi} \\
&=& -\p_\s \left[\left(  f(\hat u) - \frac{2\mu \D \hat u}{R_H} Q_{\hat f'}\right)\, \frac{\p_{\phi'}\H +\pi}{2} + \frac{2\mu}{R_u} \left( \F_f + Q_{\hat f'} \left( \frac{R_v \Delta \hat v}{R_H} -2t\right)  \right)\,\frac{\p_{\phi'}\H -\pi}{2} \right] \nonumber 
\eea
 using  the Poisson bracket of the field-dependent coordinate with the conjugate momentum $\pi$  

\be
\left\{ \frac{\tilde u}{R_u}, \pi \right\} = \p_\s \left[  \frac{2\mu \Delta \hat u}{R_H} \cdot  \frac{\p_{\phi'} \H + \pi}{2} -  \frac{2\mu}{R_u} \left( G(\tilde \s -\s) + \frac{R_v \Delta \hat v}{R_H} - \Delta \hat{\tilde u} -2t\right) \frac{\p_{\phi'} \H -\pi}{2} \right]
\ee
These  expressions for the $\phi, \pi$ variations can be  written more compactly as

\be
\d_f \phi = -\xi^t \p_\pi \H - \xi^\s \phi' \;, \;\;\;\;\; \d_f \pi = - \p_\s (\xi^t \p_{\phi'} \H + \xi^\s \pi) 
\ee
where the components $\xi^\a$ of the diffeomorphism associated to these transformations can be read off explicitly from \eqref{varphi} and \eqref{varpi}. Given that, also, $\d_f \phi = - \p_\pi Q_f$ and $\d_f \pi = - \p_\s \p_{\phi'} Q_f$, we can  identify
\be
\xi^t = \frac{\p Q_f}{\p\H} \;, \;\;\;\;\; \xi^\s = \frac{\p Q_f}{\p \P}
\ee
where the derivative uses the fact that the field-dependent coordinates in $\TT$ - deformed CFTs are entirely constructed from the Hamiltonian and momentum densities.

\subsection{From Hamiltonian to Lagrangian transformations}

We would now like to translate these symmetry transformations to the Lagrangian formalism. The explicit components of the (left-moving) diffeomorphism we have found are


\be
\xi^U =  f(\hat u) - \frac{2\mu Q_{\hat f'}}{R_H} (\hat u - U) \;, \;\;\;\;\; \xi^V = -\frac{2\mu}{R_u} (\F_f- Q_{\hat f'} U)- \frac{2\mu Q_{\hat f'}}{R_u R_H} (R_v \hat v +2\mu H_R V ) \label{newdiff}
\ee
where we used the definition \eqref{eq:uvDef} of the field-dependent coordinates and  $\F_f$ is given in \eqref{Ff}. The derivatives of this function are
\be
\p_\s \F_f = - \hat f'(\hat u) \H_L + Q_{\hat f'} \;, \;\;\;\;\; \frac{d \F_f}{dt} = \p_t \F_f - \{ H, \F_f\} =  - \hat f' (\hat u) \H_L \frac{1+2\mu \P}{1+2\mu \H} + Q_{\hat f'} 
\ee
Note that we have used the on-shell expression for the time derivative, as only then need the Hamiltonian and Lagrangian symmetry transformations  to agree\footnote{This fact can already be seen in the case of a two-dimensional free scalar transforming under e.g. left conformal transformations, under which $\d \phi = - f(U) (\pi+\phi')$, $\d \pi = - \p_{\s} [f(U) (\pi+\phi')]$. Even though $\pi = \dot \phi$ only using the definition of the velocity, we note that $\d_f \pi = \d_f \dot \phi$ only upon using \emph{both} Hamiltonian equations of motion.}. 
Replacing $\H_{L,R}$ by their Lagrangian expressions \eqref{HLRlagr} and using the expressions for the stress tensor components given e.g. in \cite{Guica:2020uhm}, we deduce that in the Lagrangian formalism


\be
\p_U \F_f = 
- \hat f' T_{UU}+ Q_{\hat f'}\;, \;\;\;\;\;\;\;\;  \p_V \F_f = 
 \hat f' T_{UV}
\ee
Note that, on-shell, the resulting diffeomorphisms  precisely agree with \eqref{holodiffs}, including the winding terms, upon taking into account the fact that $Q_{f'} = \frac{1}{R_u} Q_{\hat f'}$. 

 The variation of the action under this diffeomorphism is given by \eqref{varactdiff}. Assuming 
 for simplicity that $Q_{f'}$ is constant, we compute
%
%
\be
T_{\a\b}\, \p^\a \xi^\b =-2\left(  f'(u)- \frac{2\mu Q_{ f'}}{R_H}\right)  T_{UV} 
\ee
and thus  the action is not invariant. However, since the overall non-invariance of the action is proportional to the trace of the stress tensor, one immediately notes that it can be absorbed by a Weyl transformation $g_{\mu\nu} \r \Omega^2 g_{\mu\nu}$, with

\be
\Omega^2 = 1+ \left( f' - \frac{2\mu Q_{f'}}{ R_H}\right) \label{weyl}
\ee
where $f$ is assumed to be infinitesimal. 
The last term precisely agrees with the radial component of the  diffeomorphism  \eqref{holodiffs} in the dual space-time, which is known to implement  Weyl transformations in the boundary theory.


Thus, we find that the symmetries in the Lagrangian formalism that correspond to those generated by $Q_f, \bar Q_{\bar f}$ are given by the field-dependent coordinate transformation \eqref{newdiff}, accompanied by a Weyl rescaling \eqref{weyl} of the metric. 
 These symmetry transformations are in perfect agreement with the results of the holographic analyses. They also turn out to lead to well-defined charge variations, directly in the Lagrangian formalism. Note that from the point of view of the Lagrangian variations, the charges $Q_{f'}$ that enter the diffeomorphism much be taken to be space-time constants, e.g. by projecting onto the temporal zero mode. This is related to the fact that the Hamiltonian and Lagrangian symmetries only need to agree on-shell. 
%
%

The Noether current associated with these symmetries consists of the standard $\xi^\a T_{\a\b}$ term, but also receives contributions from the Weyl transformation. Denoting the scale factor of the metric as $\psi$, its variation found above takes the form 

\be
\d \psi = R^{\a}{}_{\b} \p_\a \xi^\b  = f' - \frac{2\mu Q_{f'}}{ R_H} \label{scalevar}
\ee
for some tensor $R_{\a\b}$, not necessarily symmetric. Taking into account  the contribution of this term to the variation of the action, we find that, on-shell, we should have 

\be
 \p_\a (T^{\a}_\b \xi^\b) + 2 T_{UV}\d\psi  = \p_\a J^\a
\ee
%
where $J^\a$ is the Noether current associated to the full symmetry. Plugging in the expression for $\d \psi$ in terms of $R_{\a\b}$, this can happen if the resulting term $T_{UV} R^{\a}{}_\b$ is conserved on-shell, as is the case if it is proportional to the stress tensor. The choice  $R_{\a U} =0$ and $T_{UV} R_{\a V} =-\frac{1}{2} T_{\a V} $ does precisely the job, where the coefficient is fixed by \eqref{scalevar}. The resulting Noether current is
%

\be
J_\a = (T_{\a\b} +2  T_{UV} R_{\a\b}) \xi^\b = T_{\a U} \xi^U 
\ee
in perfect agreement with the result of  the Hamiltonian formalism and our previous expressions for the charges.

\section{Quantum aspects of the symmetries}
\label{sec:Quant}

What makes the analysis of \emph{classical} $T\bar T$ - deformed CFTs particularly simple is the fact that the $\TT$ ``operator'' is exactly known. This is no longer the case in the quantum theory, where derivative ambiguities in the definition of the full $\TT$ operator (which, in our notation, would correspond to the difference between the operator $\p_\mu \hat \H$ and the point-split expression \eqref{psplit}) become important. These corrections are nevertheless somewhat difficult to compute, and it is not  clear whether a well-defined prescription exists for fixing all the ambiguities that arise \cite{Rosenhaus:2019utc}.

In this section, we study  quantum $T\bar T$ - deformed CFTs to leading order in the $\TT$ coupling, and check whether  the relation between $\p_\mu H$ and the integrated point-split operator \eqref{psplit}, as well as the trace relation \eqref{eq:trEq} receive quantum corrections to this order. We then discuss how these quantum corrections are expected to propagate to the flow operator and the conserved charges, as well as to the relation between the flowed and the ``unrescaled'' symmetry generators.

\subsection{Perturbative corrections to the \TT operator \label{pertcorr}}

In section \ref{hamform}, we have discussed how, classically, the $\TT$ Hamiltonian density is entirely fixed by the commutation relations of the translational currents, themselves dictated by the conservation equation and the assumption that the theory only depends on first time derivatives. The $\TT$ flow equation for the Hamiltonian density was simply consistent with this solution.

This suggests a way of proceeding in the quantum case, namely by fixing the Hamiltonian density via the current commutation relations. This is guaranteed to yield an expression for $\hat \H(\mu)$, though not necessarily an unambiguous one. The $\mu$ derivative of this expression may or may not agree with the point-split $\TT$ operator \eqref{psplit}, and one may also explicitly check whether the local trace relation \eqref{eq:trEq} is being satisfied. 

We will be denoting the Fourier modes of $\H, \P$ and $T_{\s\s}$ as $H_m, P_m$ and respectively $K_m$.  The classical commutators \eqref{pbcurr} read, in this notation

\be
[H_m, H_n] = [P_m,P_n] = (m-n) P_{m+n}  \;, \;\;\;\;\; [H_m, P_n] = m H_{m+n} - n K_{m+n} \label{clscomm}
\ee
In a CFT$_2$ one has, in terms of the more standard generators $L_m, \bar L_m$ 

\be
H_m^{(0)} = K_m^{(0)} = L_m + \bar L_{-m}\;, \;\;\;\;P_m^{(0)} = L_m - \bar L_{-m}
\ee
and one can easily check the  commutation relations are satisfied also quantum-mechanically if one adds $\frac{c}6 m^3 \d_{m+n}$ to  the right-hand-side of the second equation.  More generally, the first two equations in \eqref{clscomm} must be satisfied for either $m$ or $n$ equal to zero, whereas the second equation must be satisfied when $m=0$, as in these cases  correspond to the current conservation equations.

According to our discussion in section \ref{flopcls}, quantum-mechanically the components of the stress tensor satisfy
\be
\p_\mu H_m = - (\O_{T\bar T})_m + m V_m  \;, \;\;\;\;\; K_m = H_m -2 \mu (\O_{T\bar T})_m +m W_m \label{qreltocheck}
\ee
where $(\O_{T\bar T})_m$ is the $m^{th}$ Fourier mode of the point-split expression \eqref{psplit}

\be
(\O_{T\bar T})_m =  \sum_k \frac{1}{2} (H_k K_{m-k} + K_k H_{m-k}) - P_k P_{m-k} \label{pspfour}
\ee
and  $V_m, W_m$ represent  total derivative ambiguities. The factor of $m$ that multiplies them is due to the fact that these ambiguities are assumed to be local, case in which a time derivative (or, a commutator with the Hamiltonian) brings down a factor of $m$.

We would now like to compute these corrections  to first order in perturbation theory. Classically, we have
\be
\H(\s)= \H^{(0)} - 4 \mu \H_L^{(0)} \H_R^{(0)} +8 \mu^2 \H^{(0)} \H_L^{(0)} \H_R^{(0)} +\O(\mu^3)\;, \;\;\;\;\; \P(\s) = \P^{(0)}\ee

\be
 \K(\s) = \H^{(0)} - 12 \mu \H_L^{(0)} \H_R^{(0)} + 10 \mu^2 \H^{(0)} (\H_0^2-\P^2) + \O(\mu^3)
\ee
We see that to first order in $\mu$ there are no ordering ambiguities at the quantum level, though we could have quantum corrections. We thus make the following Ansatz for the quantum generators up to $\O(\mu)$
\be
H_m = L_m +\bar L_{-m} -4 \mu  \sum_k L_{m+k} \bar L_k + \Delta H_m \;, \;\;\;\;\;\; P_m = L_m -\bar L_{-m} +\Delta P_m
\ee
where $L_m, \bar L_m$ are the \emph{undeformed} CFT generators and the corrections are assumed to take the general form
\be
\Delta H_m = \a_m (L_m+\bar L_{-m}) +\g_m (L_m-\bar L_{-m}) \;, \;\;\;\;\;\; \Delta P_m = \b_m (L_m-\bar L_{-m}) +\d_m (L_m+\bar L_{-m})
\ee
where the coefficients are all $\O(\mu)$. Dimensional analysis then suggests that they should be proportional to $\mu m^2$, as this would correspond to e.g. a correction
 of the form $\mu\, \p^2 \H \r \mu\,  m^2 H_m$. Note that the freedom to redefine the currents via an improvement term (which acts as $\p_\s$ on $\H$ and as $[H,\,\cdot\;]$ on $\P$) shifts $\a_m$ and $\b_m$ by the same amount, while an independent redefinition simultaneously shifts $\g_m$ and $\d_m$. 
 
 We will now compute the commutation relations of these currents, using the  commutator of the Virasoro generators on the cylinder 

\be
[L_m, L_n] = (m-n) L_{m+n} + \frac{c}{12} m^3 \d_{m+n}
\ee
where the Casimir energy has been absorbed into the definition of $L_0$. To $\O(\mu)$, we find
\bea
[H_m,H_n] 
&=& (m-n) P_{m+n} + (m-n)[\a_m+\a_n - \b_{m+n} + \frac{\mu c}{3} (m^2+n^2+mn)] \, P^{(0)}_{m+n}  \\
&& \hspace{5mm}+\; (m-n)(\g_m+\g_n - \d_{m+n}) \,  H^{(0)}_{m+n}  \nonumber \\[5pt]
\left[P_m, P_n\right] &=& (m-n) P_{m+n} + (m-n) (\b_m + \b_n -\b_{m+n}) P^{(0)}_{m+n} +(m-n) (\d_m+\d_n-\d_{m+n})  H^{(0)}_{m+n}\!\!\!\!\!\!\!\!\!  \nonumber
\eea
The terms proportional to $m^3$ and $n^3$ in the $[H_m, H_n]$ commutator are problematic, as they affect current conservation when $m$ or $n$ are zero. Consequently, they should be cancelled by an appropriate choice of $\a_m$ and $\b_m$. 
 Writing $\a_m = \a  m^2$ and $\b_m = \b m^2$, this requirement reads

\be
\a - \b = - \frac{\mu c}{3}
\ee 
Thus, we find that current conservation fixes precisely  the part of the correction to the current that is not affected by the improvement ambiguity.  
 Similarly letting $\g_m = \g m^2$, $\d_m = \d m^2$, the  commutator of two momenta requires that $\d = \g$, a relation that, again, is not affected by the improvement ambiguity. The final result for the commutators to this order is thus\footnote{The  superscript ${}^{(0)}$ is redundant, given that the coefficients of these terms are already $\O(\mu)$, but we kept it for clarity.}
\begin{eqnarray*}
[H_m,H_n] & = & (m-n) P_{m+n} +(m-n) mn \left(\frac{\mu c}{3}-2\b\right)  P^{(0)}_{m+n} - 2\g (m-n)   mn  H^{(0)}_{m+n}  \nonumber \\[4pt]
\left[ P_m,P_n \right] & = & (m-n) P_{m+n} - 2 \b (m-n) mn  P^{(0)}_{m+n}   -2 \g (m-n) mn H^{(0)}_{m+n}   \nonumber
\end{eqnarray*}
As we already mentioned, the values of $\b$ and $\g$ can be shifted by improvement terms. 
Note  that for arbitrary $m,n$ there is no choice of these constants so that the algebra is identical to the classical one. However, as we saw, the part of the current algebra that is directly related to conservation  is powerful enough,  even quantum-mechanically, to fix the ``meaningful'' part of the corrections, i.e. the one that is not affected by the current improvement ambiguity.

Let us now move on to the $[H_m,P_n]$ commutator, which should determine the corrections to $K_m$. The latter takes the form, to this order 

\be
K_m = L_m + \bar L_{-m} - 12 \mu \sum_k L_{m+k} \bar L_k + \Delta K_m \;, \;\;\;\; \Delta K_m= \e_m  (L_m + \bar L_{-m})+\theta_m  (L_m - \bar L_{-m})
\ee
Note that the current improvement terms shift  simultaneously $\e_m$ and $\b_m$, and a separate transformation shifts $\theta_m$ and $\d_m$. The commutator reads
\begin{eqnarray*}
[H_m, P_n] \!& = &\!\! m H_{m+n} - n K_{m+n} + \frac{c m^3 }{6} (1+ \a_m +\b_n) \d_{m+n} + m (\a_m + \b_n -\a_{m+n}) H_{m+n}^{(0)} +\nonumber\\
&& \hspace{-2.2cm} + \; n \left( \frac{\mu c}{3} n^2 -\a_m-\b_n + \e_{m+n}\right)H_{m+n}^{(0)} +m (\g_m +\d_n-\g_{m+n}) P_{m+n}^{(0)} + n (\theta_{m+n} -\g_m -\d_n) P_{m+n}^{(0)} \nonumber
\end{eqnarray*}
This algebra should not be corrected for $m=0$, which represents the conservation equation. Letting again $\e_m = m^2 \e$, $\theta_m = m^2\theta$, we find that

\be
\e = \b - \frac{\mu c}{3} \;, \;\;\;\; \theta=\d
\ee
whereas previously we had obtained that $\d=\g$. Therefore, the differences of these terms, which could not be shifted away by improvement terms, are entirely fixed by the consistency of the commutation relations. 

We are now finally ready to check the relations \eqref{qreltocheck}, which only involve the zeroth order $T\bar T$ operator 
\be
(\O_{T\bar T})_m = 4 \sum_k L_{m+k} \bar L_k 
\ee
finding
\be
\p_\mu H_m + (\O_{T\bar T})_m =  \p_\mu \a_m  H^{(0)}_m  
+  \p_\mu \g_m  P_m^{(0)} \label{corrflowH}
\ee

\be
K_m-H_m +2 \mu (\O_{T\bar T})_m = (\e_m -\a_m) H_m^{(0)} + (\theta_m-\g_m) P_m^{(0)} =0
\ee
Consequently, while the flow equation for the Hamiltonian density could acquire non-trivial corrections that depend on the choice of improvement terms, the trace relation for the stress tensor density does appear to be obeyed exactly to this order, regardless of improvement ambiguities. Note that $\p_\mu H$ itself is not corrected at this order, thanks to the locality assumption on the form of the corrections that we have made. However, the first-order corrections to the currents will enter the Hamiltonian at the next order, though their contribution to the $T\bar T$ operator. The correction proportional to $\p_\mu P_m$ may also be argued to vanish at this order, as it is not consistent with parity.

One can use the quantum-corrected current components to compute the $T\bar T$ operator to $\O(\mu)$ 
and then evaluate the expectation value of the integrated operator in a deformed energy eigenstate, which is expected to be finite and to factorize.  The Fourier zero mode of the  point-split $T\bar T$ operator \eqref{pspfour} is, to this order 
\bea
(\O_{T\bar T})_0 &=& 4 \sum_m  L_m \bar L_m - 16 \mu \sum_{m,k} (L_{m+k} L_{-m} \bar L_k + L_{m+k} \bar L_m \bar L_k) + \sum_m \left( a_1 m^2 L_m L_{-m} + \right. \nonumber \\
& & \hspace{-1cm}+\; \left. a_2  m^2 L_m \bar L_m + a_3  m^2 L_{-m} \bar L_{-m} +a_4m^2 \bar L_{-m} \bar L_m \right] \label{ottbzm}
\eea
with
\be
a_1=  \a+\e +\g +\theta -\d -2\b \;, \;\;\;\;\;\; a_2= a_3=\a+\e +2\b 
\ee

\be
a_4=\a+\e -\g -\theta + \d-2\b
\ee
%
%
%
We now consider the expectation value of this operator in a  deformed energy eigenstate, taken here to be a deformed primary, for simplicity. To find this state, we need the zeroth order $\hat \chi_{T\bar T}$ operator, which satisfies 
\be
 [H,\hat \X_{T\bar T}] =  \O_{T\bar T} - diag. =  4 \sum_{m\neq 0} L_m \bar L_m
\ee
Since, to this order $H = L_0 + \bar L_0$, we find
\be
\X_{T\bar T} = -2  \sum_{m\neq 0} \frac{1}{m} L_{m} \bar L_m \;\;\;\;\; \Rightarrow \;\;\; \;\;|h_\mu\rangle = |h\rangle + 2 \mu  \sum_{m=1}^\infty \frac{1}{m} L_{-m} \bar L_{-m} | h \rangle
\ee
%
  Finally, we compute
\bea
\langle h_\mu | (\O_{T\bar T})_0 | h_\mu \rangle & = & \langle h | (\O_{T\bar T})_0 | h \rangle + 8 \mu  \langle h| \sum_m L_m \bar L_m \sum_{k=1}^\infty \frac{1}{k} L_{-k} \bar L_{-k} |h\rangle+ 8 \mu \langle h|\sum_{k=1}^\infty \frac{1}{k} L_{k} \bar L_{k} \sum_m L_m \bar L_m  |h\rangle \nonumber \\
&& \hspace{-2.9cm}= \; 4 h \bar h - 16 \mu h \bar h (h+\bar h) +
2(h+\bar h) \left(\frac{2\mu c}{3} +\a+\e -2\b \right)\sum_{1}^\infty k^3 + 2(h-\bar h) (\g+\theta-\d) \sum_{1}^\infty k^3 + \nonumber \\
&&\hspace{-2.9cm} +\; \left(\frac{\mu  c^2}{9} - \frac{c}{6} (\a+\e-2\b) \right) \sum_1^\infty k^5 \label{expvalottb}
\eea
where $h, \bar h$ denote the CFT dimensions \emph{minus} $c/24$, i.e. they correspond precisely to the left- and  right-moving undeformed energies on the cylinder. We immediately note that  $\a+\e - 2 \b = - 2 \mu c/3$ for consistency. Thus, we find that the quantum corrections to the currents are essential for obtaining a finite expectation value for the $T\bar T$ operator to this order, and moreover we
 obtain exactly the same constraint on the unambiguous coefficients as from the commutation relations. We are however left with a divergent term proportional to $\g(h-\bar h)$, that should be set to zero. The parity argument mentioned previously would set this term to zero. 

Let us note that one should be very careful in manipulating expressions such as \eqref{ottbzm}, as the terms under the double sum can be rather ambiguous. For example, one may na\"{i}vely suppose that 

\be
\sum_{k,l} L_{m+k+l} L_{-k} \bar L_l = \sum_{k,l}  L_{-k} L_{m+k+l} \bar L_l 
\ee
via a simple relabeling of the Fourier modes; however, the difference of the above terms is 

\be
\sum_{k,l} [L_{m+k+l}, L_{-k}] \bar L_l = \sum_{k,l} (m+2k + l) L_{m+l} \bar L_l - \frac{c}{12} \sum_k k^3 \bar L_{-m}
\ee
which does not appear to vanish. We conclude that when dealing with such infinite sums, one should first take care to properly define and regulate them. 

Keeping in mind this important caveat, one may try to extend the algorithm we presented to higher orders in $\mu$. Suppose that the currents $H_m, P_m$ are known 
up to  $\O(\mu^p)$ at the full quantum level.  Then, the quantum correction at  $\O(\mu^{p+1})$ is constrained by  the current commutation relations. These corrections, being most likely of  derivative form, are not expected to enter the  Hamiltonian - which is integrated -  at $\O(\mu^{p+1})$. However, they may be used to check the relation between $\p_\mu H_m$ and the $T\bar T$ operator at $\O(\mu^p)$, as well as the trace relation at  $\O(\mu^{p+1})$, which only involves the  $T\bar T$ operator to $\O(\mu^p)$.  

These corrections, however, will enter non-trivially the $\TT$ operator at $\O(\mu^{p+1})$. Its expectation value at the same order will be computed using the state at order $p+1$, which involves $\hat \X_{T\bar T}$ at order $p$, in turn determined from $H$ at order $p+1$. It is  very interesting to ask whether the ambiguities that are likely left over by the current commutator analysis will enter the finite piece of the $\TT$ expectation value, which is related to the question  of whether the  $T\bar T$ deformation can be intrinsically defined at the full quantum level. This effect, if present, is however expected to only appear at $\O(\mu^2)$, which lies beyond the analysis of this paper.


\subsection{Brief comment on the quantum symmetry generators}

Suppose the expression for $\hat \X_{T\bar T}$ is known at the full quantum level. Then, the full expression for the Virasoro generators \eqref{fleqlt} can be unambiguously obtained. It is an interesting question, though, to understand the fate of the ``unrescaled'' generators at the full quantum level. 

The reason that one should be interested in this different basis of symmetry generators comes from the related study of $J\bar T$ - deformed CFTs. These correspond to an irrelevant deformation of a  two-dimensional  CFT by an  operator of dimension $(1,2)$, which results in a theory that is local and conformal on the left, and non-local only on the right \cite{Guica:2017lia}. There too, one may obtain two commuting sets of Virasoro-Kac-Moody generators by transporting the original CFT generators (which, in this case, also contain an affine $U(1)$ current) along the $J\bar T$ flow \cite{Guica:2021pzy}. However, one finds that the flowed left-moving generators are explicitly \emph{different} from the physical Virasoro generators associated to the conformal symmetry \cite{Guica:2020eab}. Consequently, the flowed generators are not the most natural set of generators with respect to which to define operator Ward identities, for example \cite{Guica:2021fkv}. In the case of $T\bar T$ - deformed CFTs, it appears  that the closest analogues of such ``physical'' operators are the ``unrescaled'' symmetry generators,  which count among them the energy and momentum operators in the theory.  

At the classical level, the relation between the flowed and the unrescaled generators is given by a simple multiplicative factor of the field-dependent radius. Quantum-mechanically,  the radius is an  operator, and does not commute with the symmetry generator under consideration. Therefore,   one needs to specify the operator ordering when writing the relation between the (completely unambiguous) Virasoro generators and the unrescaled ones. Given the symmetrized form \eqref{symmfleq} of the flow equations, a natural guess for the relation between the unrescaled ($L_m$) and flowed ($\widetilde L_m$) generators is 
\be
\tilde L_m = L_m + \mu H_R L_m + \mu L_m H_R \label{qrellq}
\ee
which respects hermiticity and has the correct classical limit.  The non-linear algebra of the $L_m$ generators can be straightforwardly computed using the quantum commutation relations with the Hamiltonian
\be
[L_m, H] = \a_m(H,P) L_m \;, \;\;\;\; [L_m, P] = m \hbar L_m \label{qcomm}
\ee
where
\be
\a_m (H,P) = \frac{1}{2\mu} \left(\sqrt{(1+2\mu H)^2 + 4 \mu m \hbar (1+2\mu P) + 4 \mu^2 m^2 \hbar^2} - (1+2\mu H) \right) \label{alpham}
\ee
that were worked out in \cite{Guica:2021pzy} using the exact quantum relation between $\tilde L_0$ and $H,P$. The result is  somewhat complicated, as the algebra receives corrections to all orders in $\hbar$, but not particularly  illuminating, reason for which we do not spell it out. It is worth mentioning that, unlike in the case of $J\bar T$ - deformed CFTs, the algebra of $L_{0, \pm 1}$ does close, thus forming a non-linear subalgebra of the infinite-dimensional one.  
The algebra of the right-moving generators, as well as the mixed left-right commutators can be worked out using the right-moving analogue of $\a_m$, which is simply obtained via the replacement $P \r -P$. Note also that the $\hbar \r 0$ limit of the above expression for $\a_m$ precisely agrees with the classical Poisson brackets \eqref{clscommH}.

The main problem with the proposal \eqref{qrellq} is that there is nothing preventing us from adding further terms proportional to $m$ and the conserved charges to it; each such modification will alter the non-linear algebra of the $L_m$ in a complicated, but entirely predictable way. One thus needs an independent definition of the operators $L_m$ to fix the relationship between the two. 

One attempted definition would be to study the quantum version of the flow equation \eqref{eq:HLRFlow}, and demand that all the terms that do not resum into the field-dependent coordinate be considered as contributors to the relationship between $\tilde L_m$ and $L_m$. These would of course include all the terms that do not contain a $\s$ derivative in the quantum analogue of \eqref{eq:HLRFlow}, which already lead to \eqref{qrellq}; however, depending on how the formal series in the field-dependent coordinate is to be defined, there could also be derivative terms that do not fit   into the definition of the ``quantum'' field-dependent coordinate and its various powers, which would result into corrections proportional to powers of $m$. An explicit expression for the quantum flow operator $\hat \X_{T\bar T}$ to at least a few orders in perturbation theory would  be the minimum necessary to understand  the required patten, if any. 

Another possible way of fixing these generators would be to study their action on quantum fields in the theory. Of course, they would be implementing ``operator-dependent'' coordinate transformations, a concept whose meaning is not clear \emph{a priori}; working out an explicit example, e.g. for the case of the $T\bar T$  - deformed free boson, could shed light on this interesting question. 

A final possibility is that the unrescaled generators satisfy a non-homogeneous, yet fixed flow equation with respect to $\X_{T\bar T}$, where the non-homogeneous term is known to all orders in $\hbar$, as is the case for the $T\bar T$ Hamiltonian. If such a definition is possible, then one should of course understand the reason behind  it. 
%
%

\section{Conclusions}

In this article, we have derived the classical symmetries of $T\bar T$ - deformed CFTs, taking  the well-defined quantum generators as a starting point. The classical generators implement field-dependent coordinate transformations, where the field-dependent coordinate emerges from the classical limit of the $T\bar T$ flow and is entirely unambiguous. We then translated our results to the Lagrangian framework, finding a different   set of transformations that leave the action invariant than were previously proposed in \cite{Guica:2020uhm}, which now consist of a field-dependent coordinate transformation accompanied by a Weyl rescaling of the flat background metric. These transformations precisely agree with those found in the holographic analyses of \cite{Guica:2019nzm,Georgescu:2022iyx}. We have thus succeeded in providing a fully consistent picture of the classical symmetries of $T\bar T$ - deformed CFTs, from several different perspectives. 

A natural generalization of our work would be to understand these symmetries at the quantum level. The flow-equation-based approach presented in this article would require an explicit understanding of the flow operator at the quantum level. We have hinted at an approach based on the consistency of the current commutators that may at least partly fix the ambiguities in its construction; whether this is able to completely fix the $T\bar T$ Hamiltonian is a very interesting question that we leave to future work.

Another important direction is to understand the action of these symmetries on the operators in the theory, appropriately defined. An analogy with the related case of $J\bar T$ - deformed CFTs suggests that the ``unrescaled'' generators may play an important role in this endeavour. A careful analysis of e.g. the $T\bar T$ - deformed free boson may shed light on this interesting question, and on whether this operator basis may have a  special physical significance.  

Understanding the full physical implications of these infinitely-extended non-local symmetries may have important consequences for the various applications of the $T\bar T$ deformation which, as mentioned in the introduction, range from the QCD string to non-AdS holography. In the latter case, understanding  how to define operators
 in (single-trace) $T\bar T$ - deformed CFTs and  compute their correlation functions  may provide a starting point for precision holography in the asymptotically linear dilaton spacetime,  which may shed light on how  holography works in our own universe. 

\subsection*{Acknowledgements}

The authors are grateful to Victor Gorbenko, Per Kraus and Sylvain Ribault for interesting discussions.  The research of MG and IT was supported in part by  the ERC Starting Grant 679278 Emergent-BH. The research of RM is supported in part by the National Science Foundation under research grant PHY-19-14412.

\begin{appendix}

\section{List of useful Poisson brackets}\label{appA}

The current commutators in classical $T\bar T$ - deformed CFTs are

 \be
\{ \H, \tilde \H\} = \{ \P, \tilde \P\} = (\P+\tilde \P) \d'(\s-\tilde \s) \;, \;\;\;\;\;\;\; \{ \H, \tilde \P\} = [\tilde \H + T_{\s\s}]\d'(\s-\tilde \s)
\ee
where
\be
 T_{\s\s} = \frac{\H+2\mu \P^2}{1+2\mu \H}
\ee
It is useful to define the left/right Hamiltonian currents
\be
\H_{L,R} = \frac{\H \pm \P}{2} 
\ee
Their commutation relations are
\be
\{ \H, \tilde \H_L\} = (\tilde \H_L + \frac{1}{2} \G_+) \d'(\s-\tilde \s)\;, \;\;\;\;\;\;\;\;\; 
\{ \H, \tilde \H_R\} = - (\tilde \H_R + \frac{1}{2} \G_-) \d'(\s-\tilde \s) \label{commHHLR}
\ee
where
\be 
\Gamma_\pm \equiv T_{\s\s} \pm T_{\s t} = \frac{(\H \pm \P)(1\pm 2\mu \P)}{1+2\mu \H}
\ee
This implies that
\be
\{H,\H_L\}=  - \frac{1}{2} \G_+'\;, \;\;\;\;\;\; \{H,\H_R\}= \frac{1}{2} \G_-'
\ee
We also note that 

\be
\{ \P, \tilde \H_L \} = (\H_L + \frac{1}{2} \tilde \Gamma_+) \d' \;, \;\;\;\;\;\; \{ \P, \tilde \H_R\} = (\H_R + \frac{1}{2} \tilde \Gamma_-) \d'
\ee
From here, we can easily obtain the commutation relations of the quantities $ \chi_{L,R}$, defined via \eqref{defchiLR}. For example, their commutator with $H$, which is needed for the conservation equation, is given by integrating \eqref{commHHLR} and then subtracting its zero mode

\be
\{ H,  \chi_L \} = - \frac{1}{2} \G_+ + \frac{1}{2} \int d\s \, \G_+ \;, \;\;\;\;\; \left\{H,{\chi}_R\right\}=\frac{1}{2}\Gamma_{-}-\frac{1}{2}\int\Gamma_{-}
\ee
The commutator of the flow operator with the Hamiltonian is
\begin{align}
			\{ \cX_\TT, H \}
			&
			= -\frac{R}{R_H} \oint d \s d \ts \, G(\s - \ts) \left( \{ \H, H \} \tilde \P + \H \{ \tilde \P, H \} \right)
			\nonumber \\
			&
			= -\frac{R}{R_H} \oint d \s d  \ts \, G(\s - \ts) \left( \P' \tilde \P + \H \tilde \cK' \right)
			 \nonumber \\
			&
			= -\frac{R}{R_H} \oint d \s \left( \H \cK - \P^2 \right) + \frac{H K - P^2}{R_H}
		\end{align}
		where the last line follows from integration by parts and the definition \eqref{eq:GDef} of $G$.
		We now use the integral of the trace equation \eqref{eq:trEq} to find
		\begin{align}
			K &= H - 2\mu \oint d \s (\H \cK - \P^2)
		\end{align}
		which we substitute to get
		\begin{align}
			\{ \cX_\TT, H \}
			&
			= -\oint d \s (\H \cK - \P^2) + \frac{H^2 - P^2}{R_H}
		\end{align}
The Poisson brackets of the $\chi_{L,R}$ with the currents are
	\begin{align}
			\label{eq:PBchiL}
				\{ \chi_L, \tilde \H_L \} &= (\tilde \H_L + \tilde \cK_L) \left( \d(\s - \ts) - \frac1R \right) - \frac12 G(\s - \ts) (\tilde \H'_L + \tilde \cK'_L)
				\ ,
				\nonumber \\
				\{ \chi_R, \tilde \H_R \} &= -(\tilde \H_R + \tilde \cK_R) \left( \d(\s - \ts) - \frac1R \right) + \frac12 G(\s - \ts) (\tilde \H'_R + \tilde \cK'_R)
				\ ,
				\nonumber \\
				\{ \chi_L, \tilde \H_R \} &= \frac12 G(\s - \ts) (\tilde \cK'_R - \tilde \H'_R) = \frac14 G(\s - \ts) (\tilde \cK' - \tilde \H')
				\ , \nonumber \\
				\{ \chi_R, \tilde \H_L \} &= \frac12 G(\s - \ts) (\tilde \H'_L - \tilde \cK'_L) =\frac14  G(\s - \ts) (\tilde \H' - \tilde \cK')
			\end{align}

%

\end{appendix}

\bibliographystyle{utphys}
\bibliography{TTbsymm}

\end{document}